\documentclass[12pt]{iopart}

\newcommand{\beq}{\begin{equation}}
\newcommand{\eeq}{\end{equation}}
\newcommand{\beqa}{\begin{eqnarray}}
\newcommand{\eeqa}{\end{eqnarray}}
\newcommand{\bra}[1]{\mbox{$\langle #1|$}}
\newcommand{\ket}[1]{\mbox{$|#1\rangle$}}
\usepackage{epsfig}

\begin{document}

\title[Eed M.\ Darwish, Spin observables for $d(\gamma,\pi)NN$ in the
$\Delta$(1232)-resonance region]{Spin observables for pion photoproduction on the 
deuteron in the $\Delta$(1232)-resonance region} 

\author{Eed M.\ Darwish \footnote{{\it E-mail address:}
eeddarwish@yahoo.com}} 

\address{Physics Department, Faculty of Science, South 
  Valley University, Sohag 82524, Egypt}

\begin{abstract}
        Spin observables for the three charge states of the pion for
        the pion photoproduction reaction on the deuteron, 
        $\gamma d\to\pi NN$, with polarized photon beam and/or
        oriented deuteron target are predicted. For the beam-target
        double-spin asymmetries, it is found that only the
        longitudinal asymmetries $T_{20}^{\ell}$ and $T_{2\pm
        2}^{\ell}$ do not vanish, whereas all the circular and the
        other longitudinal asymmetries do vanish. The sensitivity of
        spin observables to the model deuteron wave function is
        investigated. It has been found that only $T_{21}$ and
        $T_{22}$ are sensitive to the model deuteron wave function, in
        particular in the case of $\pi^0$-production above the
        $\Delta$-region, and that other asymmetries are not. 
\end{abstract}

\pacs{24.70.+s, 14.20.-c, 13.60.Le, 25.20.Lj, 21.45.+v}

\submitto{\jpg}

%\maketitle

%%%%%%%%%%%%%%%%%%%%%%%%%%%%%%%%%%%%%%%%%%%%%%%%%%%%%%%%%%%%%%%%%%%%%%%%
\section{Introduction} 
\label{sec1}
Single-pion photoproduction has been a subject of extensive
theoretical and experimental investigation for many decades (for an
overview see, e.g.,~\cite{Lag81,Kru03,Bur04}). This reaction is one of the
main sources of information on nucleon structure. It allows
investigation of resonance excitations of the nucleon, especially the
$\Delta$(1232)-excitation, and their photodecay amplitudes. The pion
photoproduction amplitude on the free nucleon is used as an input when
calculating pion photoproduction from heavier nuclei.

Pion photoproduction on light nuclei is primarily motivated by the
study of the elementary neutron amplitude in the absence of a neutron
target and the investigation of medium effects. It provides an
interesting means to study nuclear structure and gives information on
pion production on off-shell nucleon, as well as on the very important
$\Delta N$-interaction in a nuclear medium. The deuteron plays an
outstanding role in the investigation of pseudoscalar meson production
in electromagnetic reactions since its structure is very well
understood in comparison to heavier nuclei.  Furthermore, the small
binding energy of nucleons in the deuteron allows one to compare the
contributions of its constituents to the electromagnetic and hadronic
reactions to those from free nucleons in order to estimate interaction
effects.

During the last decade, several experiments have been performed, e.g.,
at MAMI in Mainz, ELSA in Bonn, LEGS in Brookhaven, and JLab in
Newport News, for preparing polarized beams and targets and for
polarimeters for the polarization analysis of ejected particles
(see~\cite{Kru03,Bur04,Dre04} and references therein).  Therefore, it
appears timely to study in detail polarization observables in pion
production on the deuteron. The aim will be to see what kind of
information is buried in the various polarization observables, in
particular, what can be learned about the role of subnuclear degrees
of freedom like meson and isobar or even quark-gluon degrees of
freedom.

Polarization observables for $\pi^-$-photoproduction on the deuteron
via the reaction $d(\gamma,\pi^-)pp$ have been studied within a
diagrammatic approach~\cite{Log00}. In that work, predictions for
analyzing powers connected to beam and target polarization, and to
polarization of one of the final protons are presented. In our
previous evaluation~\cite{Dar03,Dar03+}, the energy dependence of the
three charge states of the pion for incoherent pion photoproduction on
the deuteron in the $\Delta$(1232)-resonance region has been
investigated. We have presented results for differential and total
cross sections as well as results for the beam-target spin asymmetry
which determines the Gerasimov-Drell-Hearn (GDH) sum rule. Most
recently, we have predicted results for the $\pi$-meson spectra,
single- and double-spin asymmetries for incoherent pion
photoproduction on the deuteron in the $\Delta$(1232)-resonance
region~\cite{Dar04,Dar04new}. In particular, we have studied in detail
the interference of the nonresonant background amplitudes with the
dominant $\Delta$-excitation amplitude. We found that interference of
Born terms and the $\Delta$(1232)-contribution plays a significant
role in the calculations.  The vector target asymmetry $T_{11}$ has
been found to be very sensitive to this interference.

In this paper we study several polarization observables of photon and
deuteron target for the three charge states of the pion in
photoproduction of $\pi$-mesons on the deuteron in the
$\Delta$-resonance region with special emphasis on double-spin
asymmetries. We will also compare our predictions for the linear
photon asymmetry with recent experimental data from LEGS~\cite{Luc01}.
Furthermore, we will discuss the dependence of our results for single-
and double-polarization observables on the model deuteron wave
function. Moreover, the complete formal expressions of polarization
observables for the reaction $\vec\gamma\vec d\to\pi NN$ will be given
in this paper.  The particular interest in the double-polarization
asymmetries for the reaction $\vec d(\vec\gamma,\pi)NN$ is based on
the fact that, a series measurements of double-polarization
asymmetries in photoreactions on the proton, neutron, and deuteron
have been carried out or planned at different laboratories (see, for
example,~\cite{Luc01,San02}). To provide information for these
experiments involving polarized incident photon beam and polarized
deuteron target, we present in this paper various spin observables.

The paper is organized as follows. In Sec.~\ref{sec2}, we introduce
the general form of the differential cross section for the reaction
$d(\gamma,\pi)NN$ and present the treatment of the transition
amplitude which based on time-ordered perturbation theory. The
complete formal expressions of polarization observables for the
$\gamma d\to\pi NN$ reaction with polarized photon beam and/or
oriented deuteron target in terms of the transition matrix elements
are given in Sec.~\ref{sec3}. Details of the actual calculations and
the results will be presented and discussed in Sec.~\ref{sec4}. The
conclusions are summarized in Sec.~\ref{sec5}. Explicit expressions
for double-spin observables are given in \ref{appendixa}.
%%%%%%%%%%%%%%%%%%%%%%%%%%%%%%%%%%%%%%%%%%%%%%%%%%%%%%%%%%%%%%%%%%%%%%%%%%
\section{Pion photoproduction on the deuteron}
\label{sec2}
Since the formalism of the incoherent pion photoproduction reaction on
the deuteron $\gamma d\to\pi NN$ has been described in details in our
previous work~\cite{Dar03,Dar04}, we will indicate here only the
necessary notations and definitions.

The general expression of the five-fold differential cross section is
given by~\cite{BjD64}
\beqa
d^5\sigma &=& (2\pi)^{-5}\delta^{4}\left( k+d-p_{1}-p_{2}-q\right)
\frac{1}{|\vec{v}_{\gamma}-\vec{v}_{d}|} \frac{1}{2}
\frac{d^{3}q}{2\omega_{\vec{q}}} \frac{d^{3}p_{1}}{E_{1}}
\frac{d^{3}p_{2}}{E_{2}}
\frac{M_{N}^{2}}{4\omega_{\gamma}E_{d}}
\nonumber \\ 
& &\times
~\frac{1}{6}\sum_{smtm_{\gamma}m_d} 
|{\mathcal M}^{(t\mu)}_{s m m_{\gamma} m_d}|^{2} \, ,
\label{gdcs}
\eeqa
where the definition of all kinematical variables and quantum numbers
are given in~\cite{Dar03}. This expression is evaluated in the
laboratory frame or deuteron rest frame (see Fig.~4 in~\cite{Dar03}).

Recalling the derivation given in~\cite{Dar03}, the differential
cross section of incoherent pion photoproduction on the deuteron takes
the form
\beqa
\frac{d^2\sigma}{d\Omega_{\pi}} = \int_0^{q_{max}} dq\int d\Omega_{p_{NN}} 
\frac{\rho_{s}}{6}\sum_{smtm_{\gamma}m_d} |{\mathcal M}^{(t\mu)}_{sm m_{\gamma}m_d}|^{2}\,,
\label{fivefold}
\eeqa
where the maximal pion momentum $q_{max}$ is determined by the
kinematics and $\rho_{s}$ is the phase space factor.

The general form of the photoproduction transition matrix is given by
\beqa
{\mathcal M}^{(t\mu)}_{sm m_{\gamma}m_d}(\vec{k},\vec{q},\vec{p_1},\vec{p_2})
 &=&  ^{(-)}\bra{\vec{q}\,\mu,\vec{p_1}\vec{p_2}\,s\,m\,t-\mu}\epsilon_{\mu}
(m_{\gamma})J^{\mu}(0)\ket{\vec{d}\,m_d\,00}\, , 
\eeqa
where $J^{\mu}(0)$ denotes the current operator and
$\epsilon_{\mu}(m_{\gamma})$ the photon polarization vector. The
electromagnetic interaction consists of the elementary production
process on one of the nucleons $T_{\pi\gamma}^{(j)}$ $(j=1,2)$ and in
principle a possible irreducible two-body production operator
$T_{\pi\gamma}^{(NN)}$. The final $\pi NN$ state is then subject to
the various hadronic two-body interactions as described by an
half-off-shell three-body scattering amplitude $T^{\pi NN}$. In the
following, we will neglect the electromagnetic two-body production
$T_{\pi\gamma}^{(NN)}$ and the outgoing $\pi NN$ scattering state is
approximated by the free $\pi NN$ plane wave, i.e.,
\beqa
\ket{\vec{q}\,\mu,\vec{p_1}\vec{p_2}\,s\,m\,t-\mu}^{(-)}=
\ket{\vec{q}\,\mu,\vec{p_1}\vec{p_2}\,s\,m\,t-\mu}\,.
\eeqa
This means, we include only the pure plane wave impulse approximation
(IA), which is defined by the electromagnetic pion production on one
of the nucleons alone, while a more realistic treatment including
final-state interaction (FSI) as well as two-body effects will be
reported in a forthcoming paper.

The wave function of the final $NN$-state in a coupled spin-isospin
basis which satisfies the symmetry rules with respect to a permutation
of identical nucleons has the form
\beq
  |\vec{p}_{1},\vec{p}_{2},s m,t -\mu \rangle = 
  \frac{1}{\sqrt{2}}\left(
    |\vec{p}_{1}\rangle^{(1)}|\vec{p}_{2}\rangle^{(2)} - (-)^{s+t}
    |\vec{p}_{2}\rangle^{(1)}|\vec{p}_{1}\rangle^{(2)}\right)|s
  m\,,t -\mu\rangle\,,
\eeq
where the superscript indicates to which particle the ket refers. The
deuteron wave function has the form
\beq
  \langle\,\vec{p}_{1}\vec{p}_{2}, 1
  m,\,00|\,\vec{d} m_{d},00\rangle = (2\pi)^{3}
  \delta^{3}(\,
    \vec{d}-\vec{p}_{1}-\vec{p}_2 \,)
  \frac{\sqrt{2\,E_{1}E_{2}}}
  {M_{N}} \widetilde{\Psi}_{m,m_{d}}(\vec{p}_{NN})
\eeq
with
\beqa
  \widetilde{\Psi}_{m,m_{d}}(\vec{p}\,) &=&
  (2\pi)^{\frac{3}{2}}\sqrt{2E_{d}}
  \sum_{L=0,2}\sum_{m_{L}}i^{L}\,C^{L 1 1}_{m_{L} m m_{d}}\,
  u_{L}(p)Y_{Lm_{L}}(\hat{p}) \,,
\eeqa
denoting with $C^{j_1 j_2 j}_{m_1 m_2 m}$ a Clebsch-Gordan
coefficient, $u_{L}(p)$ the radial deuteron wave function and
$Y_{Lm_{L}}(\hat{p})$ a spherical harmonics.

The matrix elements are then given in the laboratory system by 
\beqa
\label{tmat_IA_lab}
  {\mathcal M}_{sm m_{\gamma}m_d}^{(t\mu)}
  (\vec k,\vec q,\vec p_1,\vec p_2) &=&
 \sqrt{2}\sum_{m^{\prime}}\langle s 
  m,\,t -\mu|\,\Big( \langle
  \vec{p}_{1}|t_{\gamma\pi}(\vec k,\vec q\,)|-\vec{p}_{2}\rangle
  \tilde{\Psi}_{m^{\prime},m_{d}}(\vec{p}_{2})  
\nonumber\\  
& & \hspace{1cm} 
-(-)^{s+t}(\vec p_1 \leftrightarrow \vec p_2) 
\Big)\,|1 m^{\prime},\,00\rangle\, ,
\eeqa
where $t_{\gamma\pi}$ denotes the elementary production amplitude 
on the free nucleon.
%%%%%%%%%%%%%%%%%%%%%%%%%%%%%%%%%%%%%%%%%%%%%%%%%%%%%%%%%%%%%%%%%%%%%%%%%
\section{Definition of polarization observables}
\label{sec3}
The most general expression for all possible polarization observables
for incoherent pion photoproduction on the deuteron is given in terms
of the transition $\mathcal M$-matrix elements by~\cite{Dar04,Aren88}
\beqa
\mathcal O & = & Tr ({\mathcal M}^{\dagger} \Omega {\mathcal M} \rho) \nonumber \\
 & = & \sum_{\stackrel{smtm_{\gamma}m_d}{s^{\prime}m^{\prime}t^{\prime}
m_{\gamma}^{\prime}m_d^{\prime}}} \int_0^{q_{max}} dq \int d\Omega_{p_{NN}}\rho_s 
\mathcal M^{(t^{\prime}\mu^{\prime})~\star}_{s^{\prime}m^{\prime},
m_{\gamma}^{\prime}m_d^{\prime}}\vec{\Omega}_{s^{\prime}m^{\prime}sm} 
\nonumber \\
& & \hspace*{3cm}\times~
\mathcal M^{(t\mu)}_{sm,m_{\gamma}m_d}
\rho^{\gamma}_{m_{\gamma}m_{\gamma}^{\prime}} \rho^{d}_{m_dm_{d}^{\prime}}\,, 
%\nonumber \\
%& & 
\eeqa
where $\rho^{\gamma}_{m_{\gamma}m_{\gamma}^{\prime}}$ and
$\rho^{d}_{m_dm_{d}^{\prime}}$ denote the density matrices of initial
photon polarization and deuteron orientation, respectively.
$\vec{\Omega}_{s^{\prime}m^{\prime}sm}$ is an operator associated with
the observable, which acts in the two-nucleon spin space.

As shown in~\cite{Dar04,Aren88} all possible polarization
observables for the reaction $d(\gamma,\pi)NN$ with polarized photon
beam and/or polarized deuteron target can be expressed in terms of the
quantities
\beqa
V_{IM}  &=& 
\frac{1}{2\sqrt{3}}~\sum_{smt,m_{\gamma}m_dm_d^{\prime}}\hspace*{-0.2cm} (-)^{1-m_d^{\prime}}  
\sqrt{2I+1} \left( \begin{array}{ccc}  1 & 1 & I \\
m_d & -m_d^{\prime} & -M \end{array} \right) 
\nonumber \\  
& &\times 
\int_0^{q_{max}}\hspace*{-0.3cm}dq \int d\Omega_{p_{NN}} \rho_s  \mathcal M^{(t\mu)~\star}_{sm,m_{\gamma}m_d}\mathcal M^{(t\mu)}_{sm,m_{\gamma}m_d^{\prime}}\,,
\label{VIM}
\eeqa
and
\beqa
W_{IM} & = &
\frac{1}{2\sqrt{3}}~\sum_{smt,m_{\gamma}m_dm_d^{\prime}}\hspace*{-0.2cm}(-)^{1-m_d^{\prime}} 
\sqrt{2I+1} \left( \begin{array}{ccc}  1 & 1 & I \\
m_d & -m_d^{\prime} & -M \end{array} \right) 
\nonumber \\ 
& & \times 
\int_0^{q_{max}}\hspace*{-0.3cm}dq \int d\Omega_{p_{NN}}\rho_s \mathcal M^{(t\mu)~\star}_{sm,m_{\gamma}m_d}\mathcal M^{(t\mu)}_{s-m,m_{\gamma}-m_d^{\prime}}\,,
\label{WIM}
\eeqa
where the Wigner $3j$-symbols are given by~\cite{Edm57}
\beqa
\left( \begin{array}{ccc}  1 & 1 & I \\
m_d & -m_d^{\prime} & -M \end{array} \right) &=&
\frac{(-)^M}{\sqrt{2I+1}}~C^{1 1 I}_{m_{d} -m_{d}^{\prime} -M}\,.
\label{3js}
\eeqa
These quantities have the symmetry properties
\beqa
V_{IM}^{\star} & = & (-1)^M ~V_{I-M}\,, \nonumber \\
W_{IM}^{\star} & = & (-1)^I ~W_{IM}\,.
\eeqa

The unpolarized differential cross section is then given by
\beqa
\frac{d^2\sigma}{d\Omega_{\pi}}  &=&  V_{00}\,.
\eeqa
A quantity of great interest is the photon asymmetry for linearly
polarized photons, which takes the form 
\beqa
\Sigma ~\frac{d^2\sigma}{d\Omega_{\pi}}  &=&  - W_{00}\,.
\eeqa
The vector target asymmetry is given by
\beqa
T_{11}~\frac{d^2\sigma}{d\Omega_{\pi}}  &=&  2~\Im m V_{11}\,.
\label{T11}
\eeqa
The tensor target asymmetries can be written as
\beqa
T_{2M} ~\frac{d^2\sigma}{d\Omega_{\pi}} & = & (2-\delta_{M0})~\Re e V_{2M}\,,~~
  M=0,1,2\,.
\eeqa
The explicit expressions for single-spin asymmetries are given in our 
previous work~\cite{Dar04}.

The photon and target double-polarization asymmetries are given by \\
(i) Circular asymmetries
\beqa
T_{1M}^c ~\frac{d^2\sigma}{d\Omega_{\pi}}  &=&  (2-\delta_{M0})~\Re e
  V_{1M}\,,~~M=0,1\,,
\label{d1}
\eeqa
\beqa
T_{2M}^c ~\frac{d^2\sigma}{d\Omega_{\pi}}  &=&  2~\Im m V_{2M}\,,~~M=0,1,2\,,
\label{d2}
\eeqa
(ii) Longitudinal asymmetries
\beqa
T_{1M}^{\ell} ~\frac{d^2\sigma}{d\Omega_{\pi}}  &=&  i ~W_{1M}\,,~~M=0,\pm 1\,,
\label{d3}
\eeqa
\beqa
T_{2M}^{\ell}~\frac{d^2\sigma}{d\Omega_{\pi}}  &=&  - W_{2M}\,,~~M=0,\pm 1,\pm 2\,.
\label{d4}
\eeqa
We list explicit expressions of all double-polarization asymmetries
for polarized photon beam and oriented deuteron target in terms of the
transition matrix elements in \ref{appendixa}.
%%%%%%%%%%%%%%%%%%%%%%%%%%%%%%%%%%%%%%%%%%%%%%%%%%%%%%%%%%%%%%%%%%%%%%%%%%
\section{Results and discussion}
\label{sec4}
In this section we present our predictions of the polarization
observables defined in Sec.~\ref{sec3}. The contribution to the pion
production amplitude in (\ref{tmat_IA_lab}) is evaluated by taking
a realistic $NN$ potential model for the deuteron wave function. In
our calculations, the wave function of the Paris potential~\cite{La+81}
has been used. For the elementary pion photoproduction operator, the
effective Lagrangian model of Schmidt {\it et al.}~\cite{ScA96} has
been considered.

The discussion of our results is divided into two parts. In order to
give a complete overview on all possible polarization observables for
the reaction $d(\gamma,\pi)NN$, we first discuss the single-spin
asymmetries $\Sigma$, $T_{11}$, $T_{20}$, $T_{21}$, and $T_{22}$ as
functions of pion angle at values for photon lab-energies different
from that presented in our previous work~\cite{Dar04}. In the second
part, we consider the double-polarization asymmetries for photon and
deuteron target as functions of emission pion angle for all the three
isospin channels of the $\vec d(\vec\gamma,\pi)NN$ reaction at nine
different values of photon lab-energy $\omega_{\gamma}=250$,
$290$, $310$, $340$, $360$, $390$, $420$, $470$, and $500$ MeV.
%%%%%%%%%%%%%%%%%%%%%%%%%%%%%%%%%%%%%%%%%%%%%%%%%%%%%%%%%%%%%%%%%%%%%%
\subsection{Single-spin asymmetries}
\label{sec41}
\begin{center}
\begin{figure}[htp]
\includegraphics[scale=0.85]{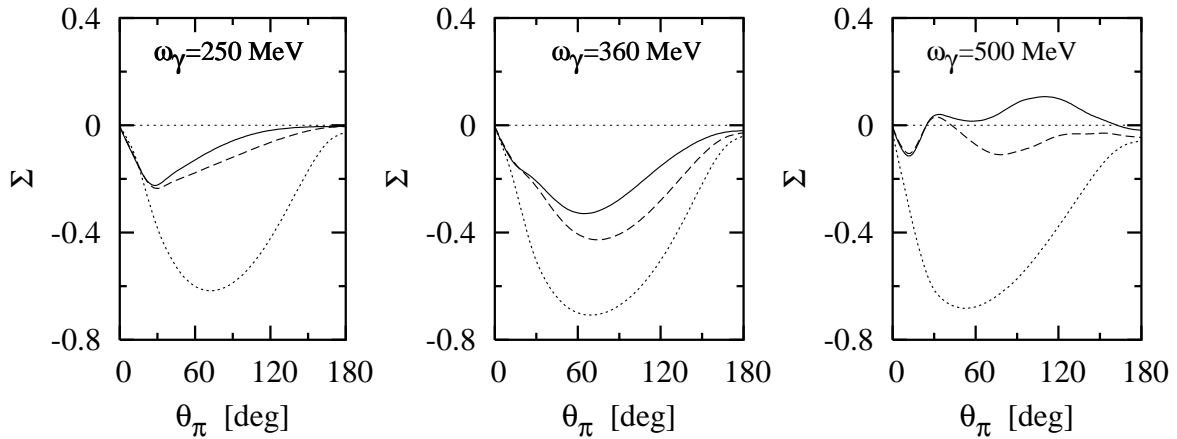}
\caption{Calculated linear photon asymmetry $\Sigma$ for the 
$d(\vec\gamma,\pi)NN$ reaction as a function of pion angle
$\theta_{\pi}$  in the laboratory frame at $\omega_{\gamma}=250$ MeV (left panel),  $360$ MeV (middle panel), and $500$ MeV
(right panel) using the deuteron wave function of the Paris
potential~\protect\cite{La+81}. The solid, dashed, and dotted curves 
correspond to $\vec\gamma d\to\pi^-pp$, $\pi^+nn$, and $\pi^0np$, 
respectively.}
\label{phasym1}
\end{figure}
\end{center}

\begin{center}
\begin{figure}[htp]
\hspace*{1cm}\includegraphics[scale=0.7]{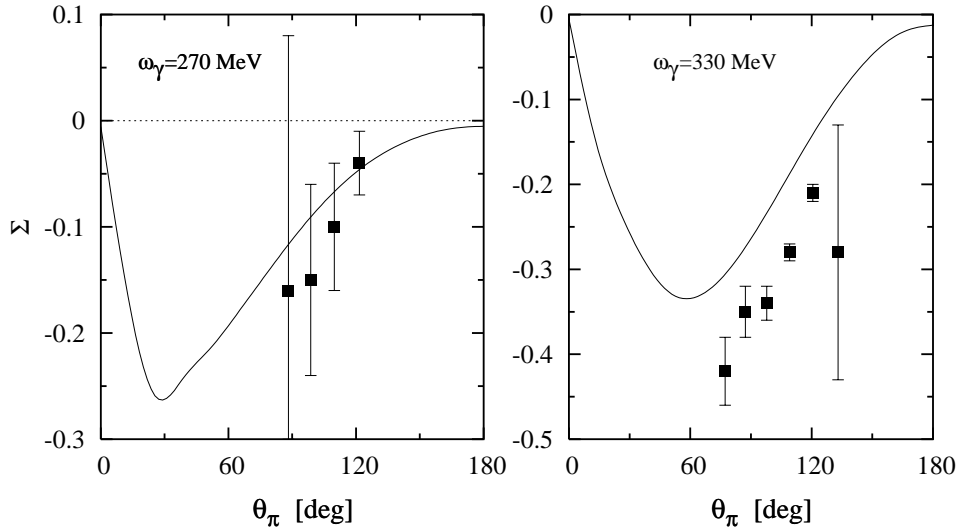}
\caption{Photon beam asymmetry $\Sigma$ for $d(\vec\gamma,\pi^-)pp$ 
  reaction in comparison with the 'preliminary' experimental data from 
  LEGS~\cite{Luc01}.}
\label{withexp}
\end{figure}
\end{center}

\begin{figure}[htp]
\includegraphics[scale=0.85]{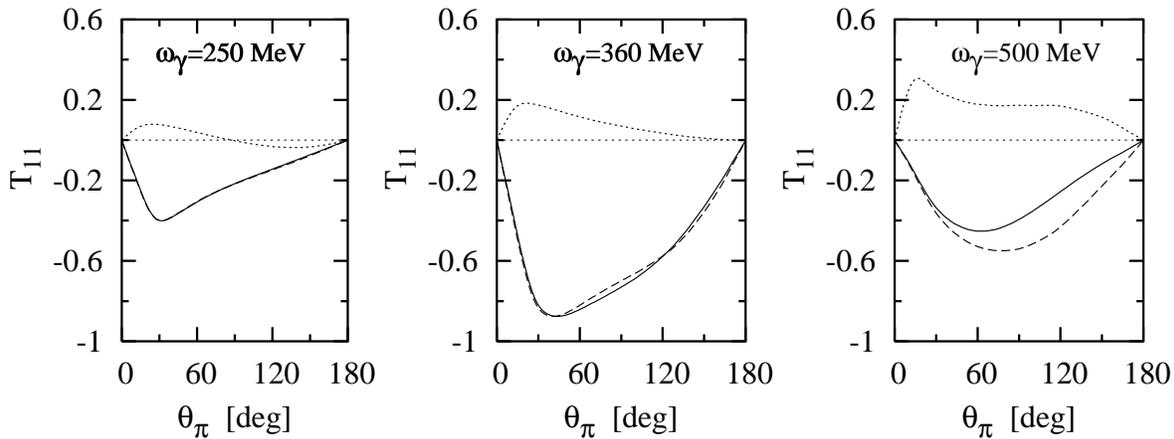}
\caption{Vector target asymmetry $T_{11}$ of $\vec d(\gamma,\pi)NN$. 
  Notation as in Fig.~\ref{phasym1}.}
\label{vtasym1}
\end{figure}

\begin{figure}[htp]
\includegraphics[scale=0.85]{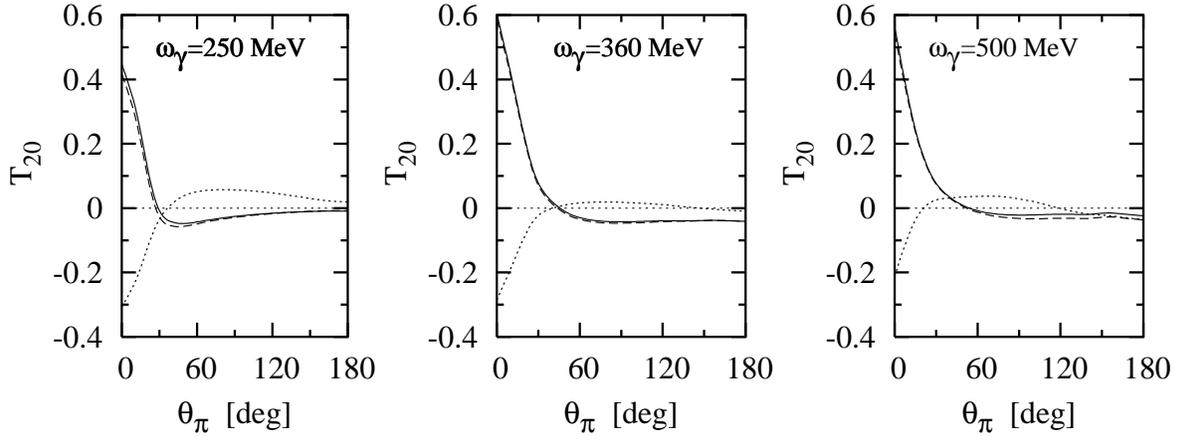}
\caption{Tensor target asymmetry $T_{20}$ of $\vec d(\gamma,\pi)NN$. 
  Notation as in Fig.~\ref{phasym1}.}
\label{ttasym201}
\end{figure}

\begin{figure}[htp]
\includegraphics[scale=0.85]{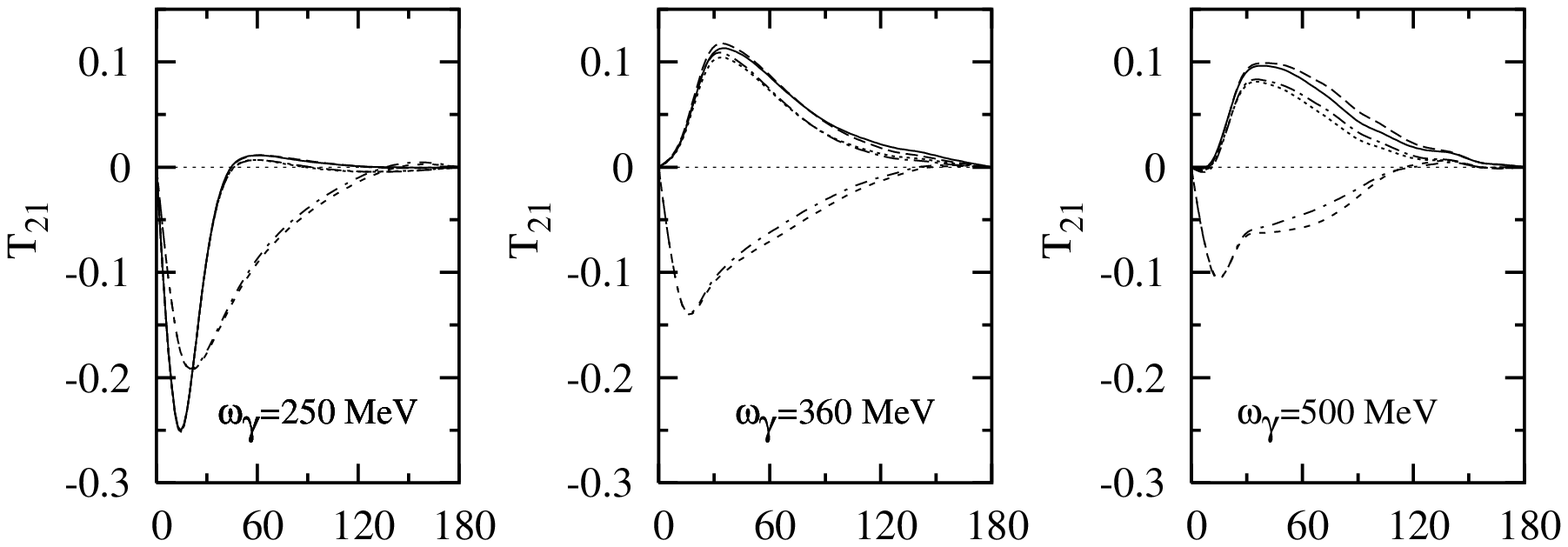}
\includegraphics[scale=0.85]{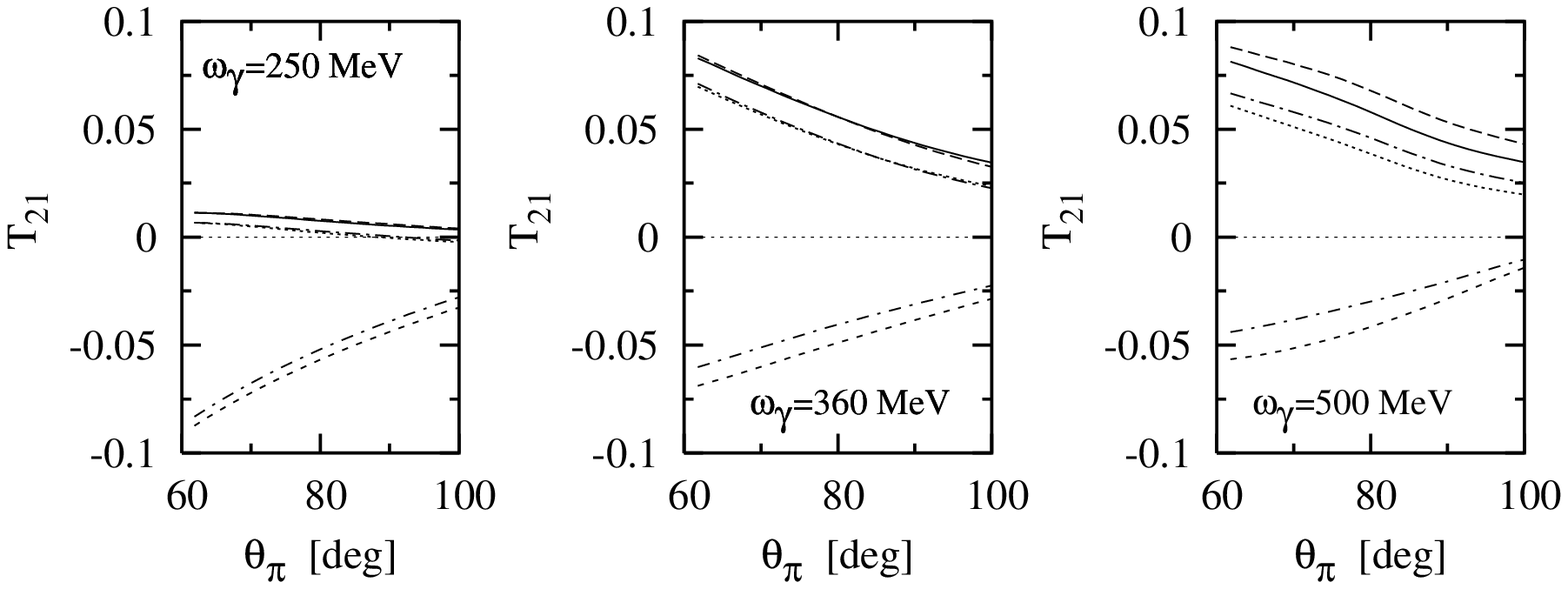}
\caption{Tensor target asymmetry $T_{21}$ of $\vec d(\gamma,\pi)NN$ at 
three different values of photon lab-energies. The solid (dotted), 
long-dashed (long-dashed-dotted), and dashed (dashed-dotted) curves 
correspond to $\gamma \vec d\to\pi^-pp$, $\pi^+nn$, and $\pi^0np$, 
respectively, using the deuteron wave function of the 
Paris~\protect\cite{La+81} (Bonn~\protect\cite{Mach}) potential model. 
Bottom panels are drawn using enlarged scales from the top ones.}
\label{ttasym211}
\end{figure}

\begin{figure}[htp]
\includegraphics[scale=0.85]{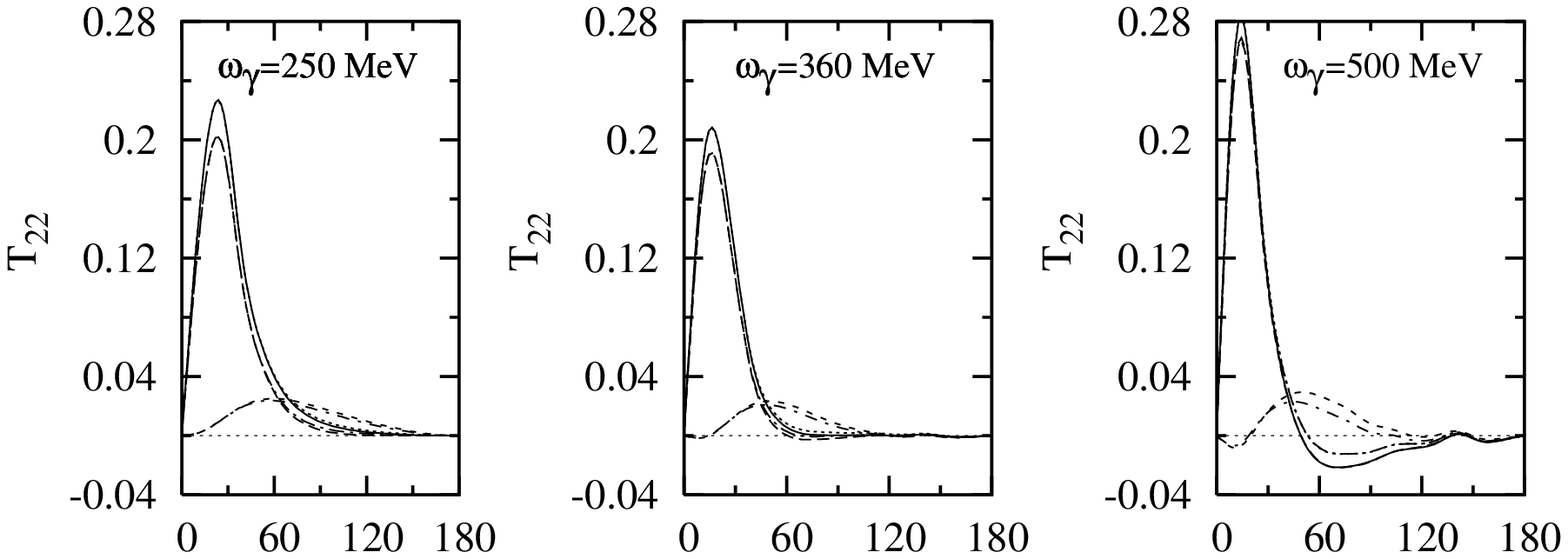}
\includegraphics[scale=0.85]{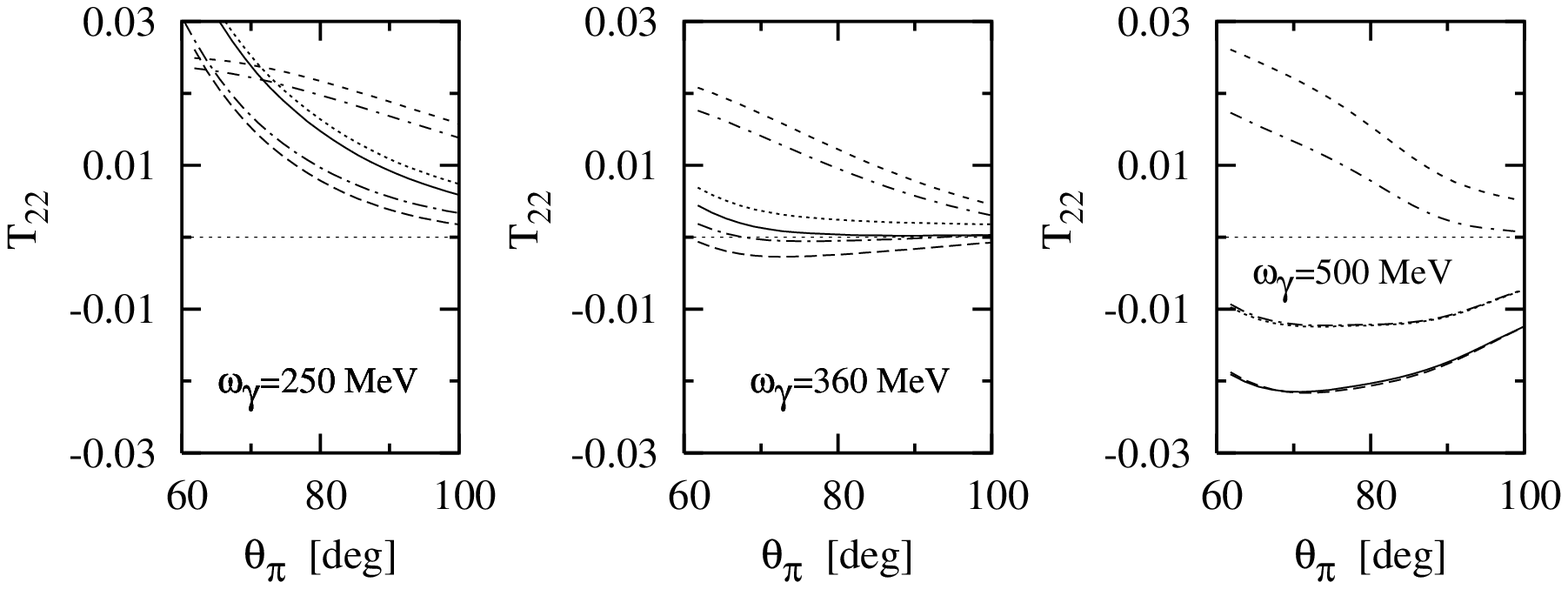}
\caption{Tensor target asymmetry $T_{22}$ of $\vec d(\gamma,\pi)NN$. 
  Notation as in Fig.~\ref{ttasym211}.}
\label{ttasym221}
\end{figure}

Let us start by discussing the results for single-spin asymmetries for
pion photoproduction on the deuteron with polarized photon beam or
polarized deuteron target as shown in Figs.~\ref{phasym1} through
\ref{ttasym221}.

The photon asymmetry $\Sigma$ for linearly polarized photons at three
values of photon lab-energy $\omega_{\gamma}=250$, $360$,
and $500$ MeV is plotted in Fig.~\ref{phasym1} as a function of pion
angle $\theta_{\pi}$ in the laboratory frame using the deuteron wave
function of the Paris potential~\cite{La+81}. The solid, dashed, and
dotted curves show the results for the $\vec\gamma d\to\pi^-pp$,
$\pi^+nn$, and $\pi^0np$ channels, respectively. In general, we see
that the linear photon asymmetry has negative values at forward and
backward emission pion angles for charged and neutral pion channels.
Only at $\omega_{\gamma}=500$ MeV, positive values are found
for $\pi^-$ production channel when $\theta_{\pi}$ changes from
$90^{\circ}$ to $150^{\circ}$. One notes qualitatively a similar
behaviour for charged pion channels whereas a different behaviour is
seen for the neutral pion channel.

At extreme forward pion angles (below $\theta_{\pi}=10^{\circ}$), one
sees that $\Sigma$ has approximately the same values for charged as
well as for neutral pion production channels, whereas a big difference
between both channels is seen when $\theta_{\pi}$ changes from
$30^{\circ}$ to $150^{\circ}$. It is also noticeable, that the
asymmetry $\Sigma$ is vanished at $\theta_{\pi}=0^{\circ}$ which is
not the case at $180^{\circ}$. 

Fig.~\ref{withexp} shows a comparison of our results for the linear
photon asymmetry $\Sigma$ with experimental data. In view of the fact
that data for $\pi^+$ and $\pi^0$ production channels are not
available, we concentrate the discussion on $\pi^-$ production, for
which we have taken the 'preliminary' experimental data from the LEGS
Spin collaboration~\cite{Luc01}. In agreement with these preliminary 
data, one can see that the predictions in the pure IA can hardly 
provide a reasonable description of the data. Major discrepancies 
are evident which
very likely come from the neglect of FSI effects.  This means that the
simple spectator approach cannot describe the experimental data.
Therefore, a careful investigation of FSI effects is necessary. An
experimental check of these predictions at wide range of pion angles
is needed. Furthermore, an independent evaluation in the framework of
effective field theory would be very interesting.

We would like to mention, that we have obtained essentially the same
results for the $\Sigma$-asymmetry if we take the deuteron wave
function of the Bonn r-space potential~\cite{Mach} instead of the
deuteron wave function of the Paris one~\cite{La+81}.

In Fig.~\ref{vtasym1} we display the results for the vector target
asymmetry $T_{11}$ as a function of pion angle $\theta_{\pi}$ at the
same values of photon lab-energies as the abovementioned case. It is
very clear that the asymmetry $T_{11}$ has different size in both
charged and neutral pion production channels, being even opposite in
phase.

For charged pion production channels we see that $T_{11}$ has negative
values, whereas positive values are observed in the case of neutral
pion production channel. Only at extreme backward pion angles, very
small negative values for $T_{11}$ is shown in the case of
$\pi^0$-production channel at $\omega_{\gamma}=250$. This is
because $T_{11}$ depends on the relative phase of the matrix elements
which can be seen from (\ref{VIM}) and (\ref{T11}). It would
vanish for a constant overall phase of the ${\mathcal M}$-matrix.  We
see also that the $T_{11}$-asymmetry is vanished at
$\theta_{\pi}=0^{\circ}$ and $180^{\circ}$. The asymmetry $T_{11}$ is
insensitive to the deuteron wave function, in agreement with the case
of photon asymmetry.

In Figs.~\ref{ttasym201}, \ref{ttasym211}, and \ref{ttasym221} we
present the results for the tensor target asymmetries $T_{20}$,
$T_{21}$, and $T_{22}$, respectively. Predictions for the three
isospin channels of $\vec d(\gamma,\pi)NN$ at three different values
of photon lab-energies are given. First we discuss the results for the
asymmetry $T_{20}$ as shown in Fig.~\ref{ttasym201}. For the reaction
$\gamma \vec d\to\pi NN$ at forward and backward emission pion angles,
the asymmetry $T_{20}$ allows one to draw specific conclusions about
details of the reaction mechanism.

Comparing with the results for linear photon and vector target
asymmetries we found that for charged pion production channels the
asymmetry $T_{20}$ has relatively large positive values at pion
forward angles (at $\theta_{\pi}<30^{\circ}$) while small negative
ones are found when $\theta_{\pi}$ changes from $30^{\circ}$ to
$180^{\circ}$. For neutral pion production channel, we see that
$T_{20}$ has negative values at forward angles and positive ones at
backward angles. Only at energies above the $\Delta$-region we observe
small negative values at extreme backward angles.

Fig.~\ref{ttasym211} displays the results for the tensor target
asymmetry $T_{21}$ as a function of pion angle $\theta_{\pi}$ using
the deuteron wave functions of the Paris and Bonn r-space potential
models. It is clear from the top panels that $T_{21}$ differs in size
between charged and neutral pion production channels. In the case of
charged pion channels, $T_{21}$ has relatively large positive values
at pion forward angles, which vanished at $\theta_{\pi}=180^{\circ}$.
Only at low energies we see that $T_{21}$ is negative at extreme
forward angles. For neutral pion channel, it is noticeable that
$T_{21}$ has negative values at forward and backward pion angles.
Furthermore, similar to the case in the vector target asymmetry
$T_{11}$, we found that $T_{21}$ is vanished at
$\theta_{\pi}=0^{\circ}$ and $\theta_{\pi}=180^{\circ}$.

As already observed in the top panels of Fig.~\ref{ttasym211}, we have
obtained a noticeable difference in the results of the asymmetry
$T_{21}$ if we take the deuteron wave function of the Bonn
potential~\cite{Mach} instead of the wave function of the Paris
one~\cite{La+81}. This difference is clear when $\theta_{\pi}$ changes
from $60^{\circ}$ to $100^{\circ}$, in particular for neutral pion
channel. At extreme forward and backward emission pion angles, we see
that this difference disappears. In order to show this difference in
more detail, we display in the bottom panels of Fig.~\ref{ttasym211}
the asymmetry $T_{21}$ using enlarged scales from the top panels.

In Fig.~\ref{ttasym221} we depict the results for the tensor target
asymmetry $T_{22}$. The same values of photon lab-energies as in the
previous figures have been used. Similar to the results of
Figs.~\ref{ttasym201} and \ref{ttasym211}, the $T_{22}$ asymmetry is
sensitive to the values of pion angle $\theta_{\pi}$. We see that, the
asymmetry $T_{22}$ has large positive values for charged pion channels
at extreme forward angles which is not the case for neutral pion
channels. Moreover, we found that $T_{22}$ is vanished at
$\theta_{\pi}=0^{\circ}$ and $180^{\circ}$.

Finally, we would like to remark that we have obtained here also
different results for the asymmetry $T_{22}$ if we take the deuteron
wave function of the Bonn potential instead of that for the Paris one
(see top panels of Fig.~\ref{ttasym221}). The bottom panels of
Fig.~\ref{ttasym221} show this difference in more detail, especially
for $\pi^0$-production channel at energies above the $\Delta$-region. 
These bottom panels are drawn using enlarged scales from the top ones. 
%%%%%%%%%%%%%%%%%%%%%%%%%%%%%%%%%%%%%%%%%%%%%%%%%%%%%%%%%%%%%%%%%%%%%%%%%%
\subsection{Beam-target double polarization asymmetries}
\label{sec42}
As already mentioned in the introduction, the main goal of this paper
is to present and discuss theoretical results for the
double-polarization asymmetries for the reaction $\vec
d(\vec\gamma,\pi)NN$.  The explicit expressions for these spin
observables are given in \ref{appendixa}. Interest in
double-polarization observables comes from the recent technical
improvements of electron accelerator facilities (such as 
LEGS~\cite{Luc01}) with both polarized beams and targets. In 
view of these recent developments, it will soon be possible to 
measure double-spin observables with precision.

First of all we would like to emphasize, that using the symmetry
relations of the Clebsch-Gordan coefficients $C^{j_1 j_2 j}_{m_1 m_2
  m}$, the asymmetries $T_{10}^c$, $T_{11}^c$, $T_{10}^{\ell}$, and
$T_{1\pm 1}^{\ell}$ (see (\ref{t10c}), (\ref{t11c}),
(\ref{t10l}), and (\ref{t1pm1l}), respectively) vanish. One should
also note, that obviously for $\Im m~|{\mathcal M}|^2 =0$, the
asymmetry $T^c_{20}$ (see (\ref{t20c})) vanishes. We found also
that the asymmetries $T^c_{21}$, $T^c_{22}$ and $T_{2\pm 2}^{\ell}$ do
vanish, whereas the spin asymmetries $T_{20}^{\ell}$ and $T_{2\pm
  2}^{\ell}$ do not. Moreover, we would like to mention that the
values for the $T_{2+2}^{\ell}$ asymmetry are found to be identical
with the values of $T_{2-2}^{\ell}$. Therefore, in what follows we
shall discuss the results for only the $T_{20}^{\ell}$ and
$T_{2+2}^{\ell}$ asymmetries.
\begin{figure}[htp]
\vspace*{1cm}
\includegraphics[scale=0.85]{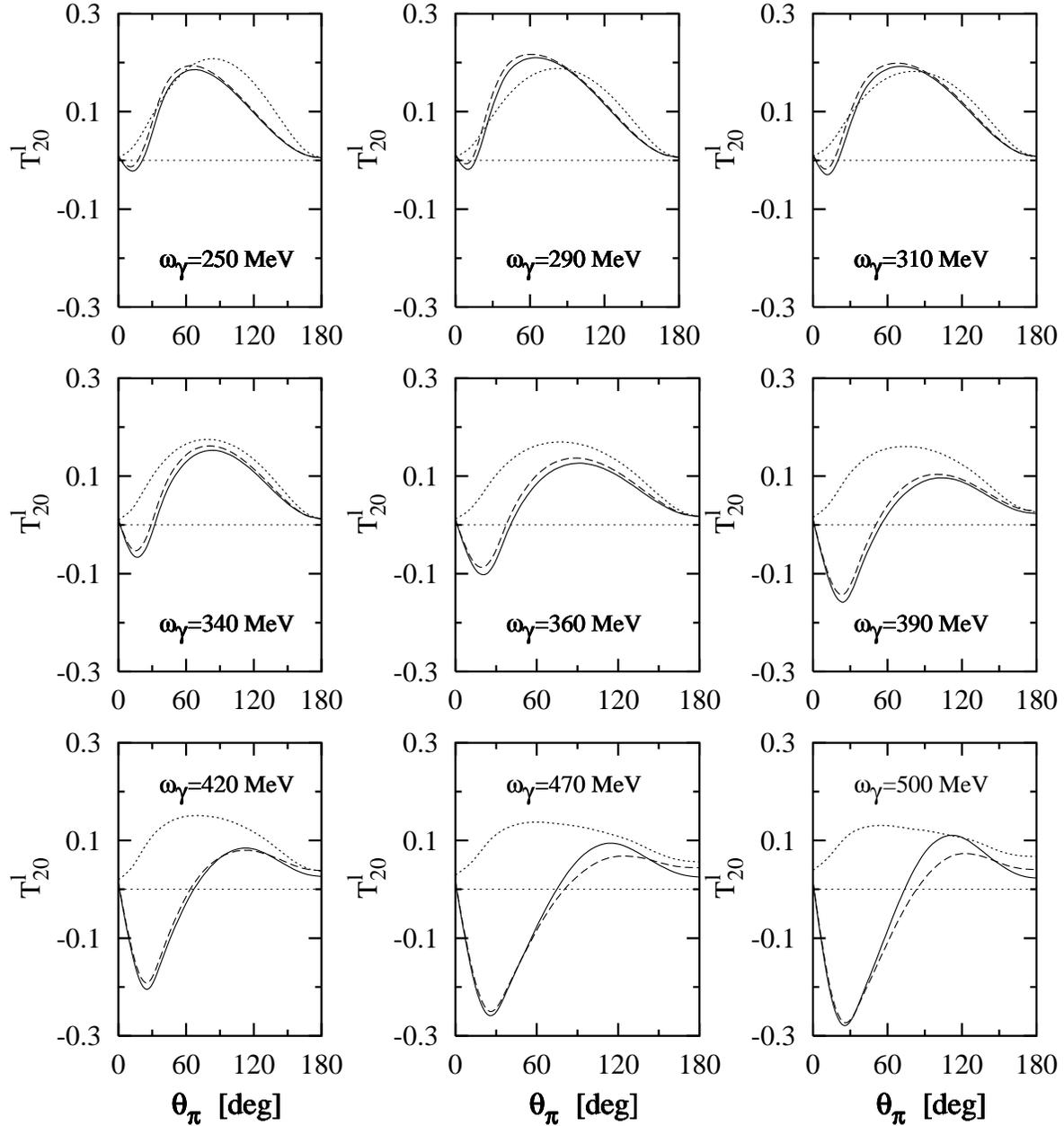}
\caption{The double-polarization asymmetry $T_{20}^{\ell}$ of 
  $\vec d(\vec\gamma,\pi)NN$ at nine different photon lab-energies. 
  Notation as in Fig.~\ref{phasym1}.}
\label{dsasymt20l}
\end{figure}

Fig.~\ref{dsasymt20l} shows the results for the longitudinal
double-spin asymmetry $T_{20}^{\ell}$ (see (\ref{t20l}) for its
definition) as a function of emission pion angle $\theta_{\pi}$ at
nine different values of photon lab-energies for the reaction
$\vec\gamma\vec d\to\pi NN$.  For $\pi^0$-production (dotted curves),
we see that $T_{20}^{\ell}$ has always positive values at all photon
energies and pion angles.  Furthermore, it is apparent that at extreme
forward and backward emission pion angles the asymmetry
$T_{20}^{\ell}$ is small in comparison with other angles.

In the case of charged pion production channels ($\pi^+$: dashed and
$\pi^-$: solid curves), we see that $T_{20}^{\ell}$ has negative
values at forward pion angles around $\theta_{\pi}=30^{\circ}$ which
is not the case at backward angles.  These negative values increase
(in absolute value decrease) with increasing the photon lab-energy.  At
extreme backward angles, we see that $T_{20}^{\ell}$ has small
positive values. The difference between the results for neutral and
charged pion channels comes from the fact that the asymmetry
$T_{20}^{\ell}$ is sensitive to the Born terms, especially at energies
above the $\Delta$-region.
\begin{figure}[htp]
\vspace*{1cm}
\includegraphics[scale=0.85]{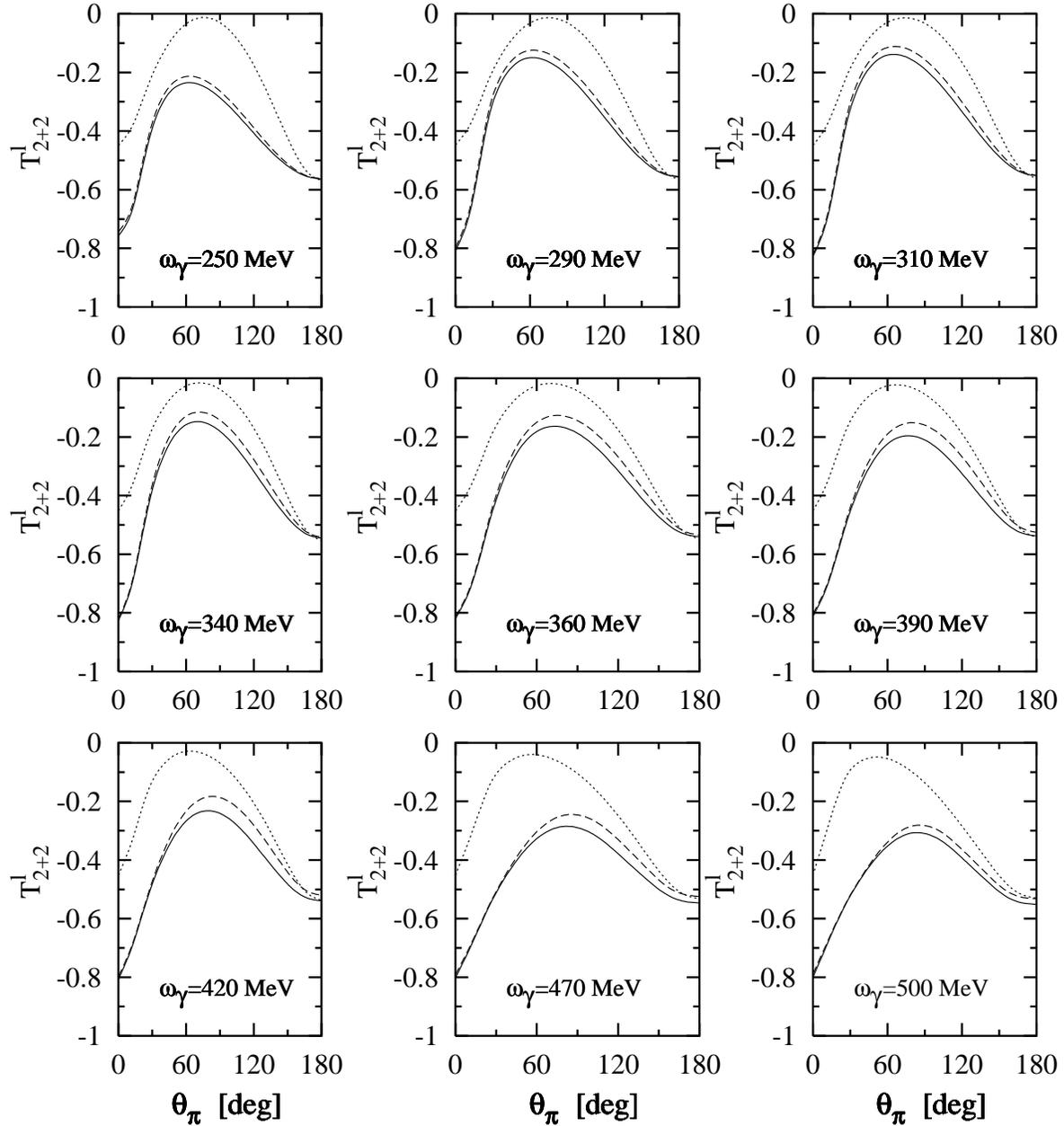}
\caption{The double-polarization asymmetry $T_{2+2}^{\ell}$ of 
  $\vec d(\vec\gamma,\pi)NN$ at nine different photon lab-energies. 
  Notation as in Fig.~\ref{phasym1}.}
\label{dsasymt2p2l}
\end{figure}

The results for the longitudinal double-polarization asymmetry
$T_{2+2}^{\ell}$ (see (\ref{t2pm2l}) for its definition) are
plotted in Fig.~\ref{dsasymt2p2l} as a function of emission pion angle
for all the three charge states of the pion for the reaction
$\vec\gamma\vec d\to\pi NN$ at the same nine values of photon
lab-energies as the abovementioned case. The solid, dashed, and dotted
curves correspond to $\vec\gamma\vec d\to\pi^-pp$, $\pi^+nn$, and
$\pi^0np$, respectively.

In general, one readily notes that the longitudinal asymmetry
$T_{2+2}^{\ell}$ has negative values.  For neutral and charged pion
production channels, it is apparent that the asymmetry
$T_{2+2}^{\ell}$ has qualitatively the same behaviour. 
We see also that the values of $T_{2+2}^{\ell}$ for $\pi^0$ channel
are greater (in absolute value smaller) than its values for
$\pi^{\pm}$ channels, in particular at $\theta_{\pi}=0^{\circ}$.
Furthermore, we noticed that the values of $T_{2+2}^{\ell}$ in the
case of $\pi^0$-production are insensitive to the photon energy and/or
pion angle, which is not the case for $\pi^{\pm}$-production channels.
It is very interesting to examine these asymmetries experimentally.

Last but not least, we would like to mention that the results for
double-spin asymmetries are insensitive to the deuteron wave function
of a particular potential model as discussed above for the single-spin
asymmetries $T_{21}$ and $T_{22}$. This means, that we have obtained 
essentially the same results for the double-spin asymmetries if we take 
the deuteron wave function of the Bonn r-space potential~\cite{Mach} 
instead of the deuteron wave function of the Paris potential 
model~\cite{La+81}.
%%%%%%%%%%%%%%%%%%%%%%%%%%%%%%%%%%%%%%%%%%%%%%%%%%%%%%%%%%%%%%%%%%%%%%%%%%
\section{Conclusions and outlook}
\label{sec5}
In this paper we have presented predictions for polarization
observables for incoherent single-pion photoproduction on the deuteron
in the $\Delta$(1232)-resonance region. Special emphasize is given for
the beam-target double-spin asymmetries, for which the explicit formal
expressions are given. The elementary $\pi$-photoproduction operator
on the free nucleon is taken in an effective Lagrangian model which
describes well the elementary reaction. For the deuteron wave
function, the realistic Paris potential model is used.

It has been found, that the beam spin asymmetry $\Sigma$ and the
target spin asymmetries $T_{11}$, $T_{20}$, $T_{21}$, and $T_{22}$ do
not vanish. For the beam-target double-spin asymmetries, it is found
that only the asymmetries $T_{20}^{\ell}$ and $T_{2\pm 2}^{\ell}$ for
longitudinal photon and deuteron target do not vanish, whereas all the
asymmetries for circular photon and deuteron target and the other
asymmetries for longitudinal photon and deuteron target do vanish. 
In comparison with experimental data for the linear photon asymmetry, 
big differences between our predictions and the 'preliminary' data from 
LEGS~\cite{Luc01} are found which very likely come from the neglect of 
FSI effects in our model.

The sensitivity of our results for single- and double-spin asymmetries
to the deuteron wave function has been investigated. For this purpose
the deuteron wave function of the Bonn potential has been used. We
found that only the tensor target asymmetries $T_{21}$ and $T_{22}$
are sensitive to the deuteron wave function of a particular $NN$
potential model, in particular in the case of neutral pion
photoproduction channel at photon energies above the
$\Delta$(1232)-resonance region, and that all other single- and
double-spin asymmetries are not. An experimental check of these 
predictions for polarization observables would provide an additional
significant test of our present understanding of low-energy behaviour
of few-body nuclei. Furthermore, in view of this low-energy property,
an independent calculations in the framework of effective field theory
would be very interesting.

In summary, we would like to point out that future improvements of the
present approach should include a more sophisticated elementary
production operator, which will allow one to extend the present
approach to higher energies, and the role of final-state interactions 
as well as two-body effects to the electromagnetic pion production
operator. This approach is necessary for the problem at hand since the
polarization observables are sensitive to the dynamical effects.
Moreover, the formalism can also be extended to investigate
polarization observables of pion electroproduction on the deuteron
where the virtual photon has more degrees of freedom than the real
one. Therefore, it can be used to explore the reaction more deeply.
%%%%%%%%%%%%%%%%%%%%%%%%%%%%%%%%%%%%%%%%%%%%%%%%%%%%%%%%%%%%%%%%%%%%%%%
\begin{appendix}
\section{Explicit expressions for double-spin asymmetries}
\label{appendixa}
In this appendix the explicit formal expressions for the various 
double-polarization asymmetries listed in (\ref{d1}) through 
(\ref{d4}) in terms of the transition ${\mathcal M}$-matrix 
elements are given.\\
(A) Asymmetries for circular photon and deuteron target
\beqa
T^c_{10} & = & \frac{\sqrt{3}}{\mathcal F} ~
\Re e\sum_{smtm_{\gamma}m_dm_d^{\prime}}
\int_{0}^{q_{\rm max}}dq~\int d\Omega_{p_{NN}}~ \rho_s~
(-)^{1-m_d^{\prime}}~C^{1 1 1}_{m_{d} -m_{d}^{\prime} 0}~
\nonumber \\ & & \hspace{5cm} \times~
{\mathcal M}^{(t\mu)~\star}_{smm_{\gamma}m_d}~
{\mathcal M}^{(t\mu)}_{smm_{\gamma}m_d^{\prime}}\nonumber \\
& =& 0\,,
\label{t10c}
\eeqa
where ${\mathcal F}$ is given by
\beqa
\label{fff}
{\mathcal F} &=& \sum_{smtm_{\gamma}m_d}\int_{0}^{q_{\rm max}}dq \int d\Omega_{p_{
NN}} ~\rho_s~ |{\mathcal M}^{(t\mu)}_{sm m_{\gamma}m_d}|^{2}\,,
\eeqa
\beqa
T^c_{11} & = & \frac{-2\sqrt{3}}{\mathcal F} ~
\Re e\sum_{smtm_{\gamma}m_dm_d^{\prime}}
\int_{0}^{q_{\rm max}}dq~\int d\Omega_{p_{NN}}~ \rho_s~
(-)^{1-m_d^{\prime}}~C^{1 1 1}_{m_{d} -m_{d}^{\prime} 1}~
\nonumber \\ & & \hspace{5cm} \times~
{\mathcal M}^{(t\mu)~\star}_{smm_{\gamma}m_d}~
{\mathcal M}^{(t\mu)}_{smm_{\gamma}m_d^{\prime}}\nonumber \\
& =& 0\,,
\label{t11c}
\eeqa
\beqa
T^c_{20} & = & \frac{\sqrt{2}}{\mathcal F} ~\Im m\sum_{smtm_{\gamma}}
\int_{0}^{q_{\rm max}}dq~\int d\Omega_{p_{NN}}~ \rho_s
~\Big[~|{\mathcal M}^{(t\mu)}_{smm_{\gamma}+1}|^2 
+ |{\mathcal M}^{(t\mu)}_{smm_{\gamma}-1}|^2
\nonumber \\ & & \hspace{5.5cm}
- 2~ |{\mathcal M}^{(t\mu)}_{smm_{\gamma}0}|^2~\Big]\nonumber \\
& =& 0\,,
\label{t20c}
\eeqa
\beqa
T^c_{21} & = & \frac{-\sqrt{6}}{\mathcal F} ~\Im m\sum_{smtm_{\gamma}} 
\int_{0}^{q_{\rm max}}dq~\int d\Omega_{p_{NN}} ~\rho_s~ 
\Big[{\mathcal M}^{(t\mu)~\star}_{smm_{\gamma}0}~
{\mathcal M}^{(t\mu)}_{smm_{\gamma}-1} 
\nonumber \\ & & \hspace{5cm}
- {\mathcal M}^{(t\mu)~\star}_{smm_{\gamma}+1}~
{\mathcal M}^{(t\mu)}_{smm_{\gamma}0}\Big]\,,
\label{t21c}
\eeqa
\beqa
T^c_{22} & = & \frac{2\sqrt{3}}{\mathcal F} ~\Im m\sum_{smtm_{\gamma}} 
\int_{0}^{q_{\rm max}}dq~\int d\Omega_{p_{NN}} ~\rho_s~ 
{\mathcal M}^{(t\mu)~\star}_{smm_{\gamma}+1}~
{\mathcal M}^{(t\mu)}_{smm_{\gamma}-1}\,.
\label{t22c}
\eeqa
(B) Asymmetries for longitudinal photon and deuteron target
\beqa
T^{\ell}_{10} & = & \frac{i\sqrt{3}}{\mathcal F} \sum_{smtm_{\gamma}m_dm_d^{\prime}}
\int_{0}^{q_{\rm max}}dq~\int d\Omega_{p_{NN}}~ \rho_s~
(-)^{1-m_d^{\prime}}~C^{1 1 1}_{m_{d} -m_{d}^{\prime} 0}~
\nonumber \\ & & \hspace{5cm} \times~
{\mathcal M}^{(t\mu)~\star}_{smm_{\gamma}m_d}~
{\mathcal M}^{(t\mu)}_{s-mm_{\gamma}-m_d^{\prime}}\nonumber \\
&=& 0\,,
\label{t10l}
\eeqa
\beqa
T^{\ell}_{1\pm 1} & = & \frac{-i\sqrt{3}}{\mathcal F} \sum_{smtm_{\gamma}m_dm_d^{\prime}}
\int_{0}^{q_{\rm max}}dq~\int d\Omega_{p_{NN}}~ \rho_s~
(-)^{1-m_d^{\prime}}~C^{1 1 1}_{m_{d} -m_{d}^{\prime} \pm 1}~
\nonumber \\ & & \hspace{5cm} \times~
{\mathcal M}^{(t\mu)~\star}_{smm_{\gamma}m_d}~
{\mathcal M}^{(t\mu)}_{s-mm_{\gamma}-m_d^{\prime}}\nonumber \\
&=& 0\,,
\label{t1pm1l}
\eeqa
\beqa
T^{\ell}_{20} & = & \frac{-1}{\sqrt{2}{\mathcal F}} \sum_{smtm_{\gamma}}
\int_{0}^{q_{\rm max}}dq~\int d\Omega_{p_{NN}}~ \rho_s~
\Big[{\mathcal M}^{(t\mu)~\star}_{smm_{\gamma}-1}~
{\mathcal M}^{(t\mu)}_{s-mm_{\gamma}+1}
\nonumber \\ & & \hspace{2cm}
+ {\mathcal M}^{(t\mu)~\star}_{smm_{\gamma}+1}~
{\mathcal M}^{(t\mu)}_{s-mm_{\gamma}-1} 
%\nonumber \\ & & \hspace{6cm}
- 2~
{\mathcal M}^{(t\mu)~\star}_{smm_{\gamma}0}~
{\mathcal M}^{(t\mu)}_{s-mm_{\gamma}0}\Big]\,, 
\label{t20l}
\eeqa
\beqa
T^{\ell}_{2\pm 1} & = & \frac{\sqrt{3}}{\sqrt{2}{\mathcal F}} \sum_{smtm_{\gamma}}
\int_{0}^{q_{\rm max}}dq~\int d\Omega_{p_{NN}}~ \rho_s~
\Big[\pm {\mathcal M}^{(t\mu)~\star}_{smm_{\gamma}0}~
{\mathcal M}^{(t\mu)}_{s-mm_{\gamma}\pm 1}
\nonumber \\ & & \hspace{5cm}
\mp {\mathcal M}^{(t\mu)~\star}_{smm_{\gamma}\pm 1}~
{\mathcal M}^{(t\mu)}_{s-mm_{\gamma}0}\Big]\,, 
\label{t2pm1l}
\eeqa
\beqa
T^{\ell}_{2\pm 2} & = & \frac{-\sqrt{3}}{\mathcal F} \sum_{smtm_{\gamma}}
\int_{0}^{q_{\rm max}}dq~\int d\Omega_{p_{NN}}~ \rho_s~
{\mathcal M}^{(t\mu)~\star}_{smm_{\gamma}\pm 1}~
{\mathcal M}^{(t\mu)}_{s-mm_{\gamma}\pm 1}\,.
\label{t2pm2l}
\eeqa
\end{appendix}
%%%%%%%%%%%%%%%%%%%%%%%%%%%%%%%%%%%%%%%%%%%%%%%%%%%%%%%%%%%%%%%%%%%%%%%
\section*{Acknowledgements}
I am gratefully acknowledge very useful discussions with Prof.\ H.\ 
Arenh\"ovel as well as the members of his work group. I am also indebted 
to Prof.\ T.-S.\ Harry Lee for fruitful discussions. I would like to 
thank Dr.\ M.\ Abdel-Aty for a careful reading of the manuscript.
%%%%%%%%%%%%%%%%%%%%%%%%%%%%%%%%%%%%%%%%%%%%%%%%%%%%%%%%%%%%%%%%%%%%%%%

\end{document}